\documentclass[%
 reprint,
 amsmath,amssymb,
 aps,
]{revtex4-1}

\usepackage{graphicx}
\usepackage{dcolumn}
\usepackage{bm}
\usepackage{amsmath}
\usepackage{breqn}

\begin{document}


\title{Conformal invariance and the Ising model on a 3 sphere in connection with the Quantum Elemental Method}
\thanks{A footnote to the article title}%

\author{Daniel Berkowitz}
 \altaffiliation{Physics Department, Yale University.}

\date{\today}

\begin{abstract}
We formulate the conformal mapping between $R^3$ and $S^3$, the 3 sphere. This mapping is applied to the critical Ising model. From this mapping, we calculate the second and fourth moments of the magnetization density, and using those quantities calculate the 4th Binder cumulant. Our calculations for the critical 3D Ising Model on a 3 sphere are done using Mathematica's Monte Carlo Integration feature. The main motivation for performing this calculation is so we can later compare it to what the Quantum Elemental Method predicts. 

\end{abstract}

\pacs{Valid PACS appear here}
\maketitle


\section{\label{sec:level1}INTRODUCTION}

In three dimensions, powerful tools have been developed to calculate conformal blocks [1-3]. Using these tools, we can realize the power that conformal invariance has on constraining quantities of interest in CFTs. Techniques for computing conformal blocks in 3D are different than those in 2D because the 3D conformal group unlike the 2D conformal group has a finite amount of generators. Thus no general closed form expressions exist in 3D for conformal blocks, instead recursion relations exist which we exploit in our calculations. 

	Advances in the Conformal Bootstrap  [4] have allowed us to calculate with precision the spin, dimension and coefficients of operators appearing in Operator Product Expansions of conformal blocks in the critical 3D Ising model  [5]  [6]. Slava Rychkov, David Simmons-Duffin and Bernardo Zan  [7] used recent advances in calculating conformal blocks and in the Conformal Bootstrap to analyze the non-gaussianity for the 3D Ising model on $R^3$ and in the process compute the numerator constrained by the conformal invariance of the four point function of the critical 3D Ising model. 
	
	In 2D Youjin Deng and Henk W. J. Blo¨te  [8] formulated the conformal mapping between an infinite plane and a spheroid. That conformal mapping was applied to the 2D critical Ising model. Through the use of continuous cluster Monte Carlo method, they calculated the second and fourth magnetization and the Bider cumulant for the 2D critical Ising model on a spheroid.  
	
	To our knowledge, no applications of conformal mappings onto curved geometries have been reported in three dimensions. In this paper, we calculate the conformal mapping of $R^3$ onto $S^3$. A 3 sphere can be constructed topologically in an analogous way to a 2 sphere. First a pair of 3 balls which are analogous to the 2 sphere but also include a degree of freedom towards the interior, of the same size are superposed onto each other, so that there boundaries match. The superposing of these two balls can be thought of as gluing together their boundaries or forming the quotient space of $S^3$. Points on the two balls that correspond to each other when glued are equivalent to each other. In analogy with the gluing together of a pair of disks with  circular $S^1$ boundaries which forms the 2 sphere, this forms a 3 sphere. Our 3 sphere is embedded in $R^4$. This extra dimension can be thought of as a continuous function of the 3 coordinates (r $\theta$, $\phi$) of the 3 ball. This function is a scalar, and we can take it to be zero along the surfaces of the 2 sphere, which corresponds to the equator of the 3 sphere. We can say that the average value of this scalar in the interior of one of the 3 balls is higher than the average value in the interior of the other 3 ball. This corresponds to the northern and southern hemispheres of our 3 sphere in our 4D Euclidean space. 
	
	We apply this mapping to the critical 3D Ising model. Using the well known bulk two and four point correlation functions in the plane and the assumption of covariance [8] of the multipoint correlations under conformal mappings, the second and the fourth moments of the magnetization density $\sigma$ on the 3 sphere can be expressed in terms of integrals. These integrals are highly non trivial and must be calculated numerically. We do this using Monte Carlo Integration on Mathematica. 
	
	Our main motivation for performing this calculation is to compare our results to a new method called the Quantum Elemental Method [9]. This numerical technique is suited for computing four point and two-point functions  of field theories on curved Riemannian manifolds. The QEM has already been successfully used to analyze the 2D critical Ising model CFT by computing its four-point function and as a result its Binder cumulant. The results of the QEM method for the Binder cumulant of the critical 2D Ising model on $S^2$ are in agreement with the results in [8]. With the QEM, Binder cumulants of the critical Ising model for a variety of Riemannian geometries  can, in principle, be calculated. Because at criticality the 3D Ising model is a CFT [4], we can compare our calculations for the Binder cumulant using Monte Carlo Integration with the results of this newly developed technique in order to advance it. Potential applications of the QEM are in calculating the quantum effects near the event horizon of a black hole in the none perturbative limit, analyzing two dimensional condensed matter systems such as graphene sheets and analyzing strongly interacting theories for BSM physics.
	
	The structure of this paper is as follows. In section 2, we give an in depth overview of the 3 sphere, including its group invariant properties. In section 3 we use the results of section II to calculate via Monte Carlo integration the Binder cumulant and magnetization densities and express our results. We then conclude this paper in section 4 where we explore future work on calculating Binder cumulants for exotic geometries so we can compare them to the cumulants the Quantum Elemental Method predicts for the Ising model at criticality on a Riemannian manifold.

\section{\label{sec:level1}CONFORMAL MAPPING OF 3 SPHERE}

In four-dimensional Cartesian coordinates (x,y,z,w) a unit 3 sphere can be defined by 
\begin{equation}
x^2+y^2+z^2+w^2=1
\end{equation}

This relation can be  parametrically satisfied in hyper spherical coordinates as the following

\begin{equation}
w=\cos{\psi}, z=\sin{\psi} \cos{\theta}, y=\sin{\psi} \sin{\theta}\sin{\phi}, x=\sin{\psi} \sin{\theta} \cos{\phi},
\end{equation}

in which $\psi$ and $\theta$ runs over the range 0 to $\pi$ and $\phi$ runs over 0 to 2$\pi$. Thus, the metric for the 3 sphere is 

\begin{equation}
ds^2=d\psi^2+\sin^2\psi (d\theta^2+\sin^2\theta d\phi^2)
\end{equation}

Before we bother calculating the Weyl factor of this metric so we can construct the four point correlation function of the critical 3D Ising on $S^3$ we will argue that this metric is conformally flat, which will allow us to map it to $R^3$. 

	All 2 manifolds are conformally flat which in essence means that a small enough patch on the manifold always exists so that a conformal transformation can map that patch to flat space [10]. In terms of an index free metric notation $g'=fg$ where g is a flat metric in a given coordinate system, f is a scalar and g' is  metric with none zero curvature, such as the metric for a 2 sphere. For 3 manifolds, a necessary and sufficient condition for conformal flatness is that its Cotton tensor vanishes. The Cotton tensor is related to the Weyl or Conformal tensor which is invariant under conformal transformations [11]. The Weyl tensor is a trace less form of the Riemann curvature tensor which measures how tidal forces distort the shape of an object in a curved space, but not how the volume of that object changes. The vanishing of the Weyl tensor for dimensions equal or greater than 4 is a necessary and sufficient for conformal flatness. The Cotton tensor for our 3 sphere can be computed in terms of the Ricci tensor, via the following equation 
	
	\begin{equation}
	C_{ijk}=\bigtriangledown_k R_{ij}-\bigtriangledown_j R_{ik}+(1/4)(\bigtriangledown_j Rg_{ik}-\bigtriangledown_k Rg_{ij})
	\end{equation}.

	Using the Mathematica package GREATER2 [12] we compute that $C_{ijk}=0$ thus the 3 sphere is conformally flat. We shall now compute its Weyl factor. 
	
	We begin by performing a coordinate transformation 
\begin{equation}
\psi=f(r)
\end{equation}
	
	where r is a conformal coordinate, plugging this into our metric (3) yields 
	
\begin{equation}
ds^2=(\frac{df}{dr})^2 dr^2+\sin^2f(r) (d\theta^2+\sin^2\theta d\phi^2)
\end{equation}
factoring out $ \frac{df}{dr}$ gives 

\begin{equation}
ds^2=(\frac{df}{dr})^2(dr^2+\frac{\sin^2f(r)}{(\frac{df}{dr})^2} (d\theta^2+\sin^2\theta d\phi^2))
\end{equation}

$(\frac{df}{dr})^2$ is our Weyl factor, in order to for this metric to be conformally flat we need this Weyl factor to be multiplied by a flat metric.\\

By examination if 
$\frac{\sin^2f(r)}{(\frac{df}{dr})^2}=r^2 $
then we will have a Weyl factor times a flat metric which in this case will be the metric for a flat space represented in 2D "spherical coordinates" . 

Using Mathematica to solve this simple ODE we get

\begin{equation}
\psi=2 \cot ^{-1}\left(\frac{1}{r}\right)=f(r).
\end{equation} 

Thus our Weyl factor is 

\begin{equation}
\Omega(r)^2=\frac{4}{\left(r^2+1\right)^2}
\end{equation}

\begin{equation}
ds^2=\frac{4}{\left(r^2+1\right)^2}(dr^2+r^2 (d\theta^2+\sin^2\theta d\phi^2))
\end{equation}

We can solve for r in terms of $\psi$ which yields $r=\tan \left(\frac{\psi }{2}\right)$. Using this we can recast our metric as

\begin{equation}
ds^2=4 \cos ^4\left(\frac{\psi }{2}\right)(dr^2+r^2 (d\theta^2+\sin^2\theta d\phi^2))
\end{equation}

where as we remind the reader $r=\tan \left(\frac{\psi }{2}\right)$. 

The coordinates to define distance for the flat portion of our metric are usual spherical coordinates

\begin{equation}
 z=r\cos\theta, y=r\sin\theta \sin\phi , x=r\sin\theta \cos\phi
\end{equation}

and we know what r is in term of $\psi$ from (9) thus we can specify any point on the surface of our 3 sphere in terms of these conformal coordinates. Even though the 3 sphere is embedded in four dimensional space, it only requires 3 coordinates to specify a point on the surface of it. These conformal coordinates are 

\begin{equation}
 z=\tan \left(\frac{\psi }{2}\right) \cos\theta, y=\tan \left(\frac{\psi }{2}\right) \sin\theta \sin\phi, x=\tan \left(\frac{\psi }{2}\right) \sin\theta \cos\phi
\end{equation}

and we will use them to construct conformal cross ratios for our four point function. 
In conjunction with the Weyl factor via conformal invariance in 3D, we can turn correlation functions in flat space to correlation functions on the curved space of $S^3$. 

	Like the 2 sphere, the 3 sphere is invariant under Special Orthogonal Group Transformations. The 2 sphere is invariant under SO(3) rotations while the 3 sphere is invariant under SO(4). SO(N) has $\frac{N(N-1)}{2}$ generators, thus the 3 sphere embedded in $R^4$ has six simple rotations associated with it. We will use this to reduce the dimension of our integrals for the 4 point and 2 point correlation function from 12 and 6 to 6 and 1 respectively. This can be seen because SO(4) is locally isomorphic to $SO(3) \otimes SO(3)$ there are six independent rotations in $R^4$, 3 for each SO(3). This can be seen by noticing the Lie Algebra of SO(4) can be represented as two copies of the Lie Algebra of SO(3).

\section{\label{sec:level1}Exact Calculation}

To  calculate our Binder cumulant we first need the magnetization densities $\langle \sigma^2 \rangle$ and $\langle \sigma^4 \rangle$.  These are given in terms of the two and four point correlation functions. The two point function for a CFT on a curved manifold is [7] 

\begin{equation}
\langle \phi(x_1)\phi(x_2)\rangle_{g_{uv}}= \frac{1}{\Omega(x_1)^\Delta} \frac{1}{\Omega(x_2)^\Delta} \langle \phi(x_1)\phi(x_2)\rangle_{flat}
\end{equation}

where we computed $\Omega(x_i)$ in (9). The 4 point function in curved space is 

\begin{equation}
\langle\phi(x_1)\phi(x_2)\phi(x_3)\phi(x_4)\rangle_{g_{uv}}= \frac{1}{\Omega(x_1)^\Delta}...\langle\phi(x_1)...\phi(x_4)\rangle_{flat}
\end{equation}

where $\Delta$ in our correlation functions is the scaling dimension of the spin fields. The scaling dimensions were computed with the Conformal Bootstrap in [13]. For the critical 3D Ising model $\Delta$ = 0.5181489. 

The two point function in flat space is 
	
\begin{equation}
\langle\phi(x_1)\phi(x_2)\rangle_{flat}=\frac{1}{x_{12}^{2\Delta}}
\end{equation}
where $x_{ij}$ =$|x_i-x_j|$. 

The four point function in flat space is 

\begin{equation}
\langle\phi(x_1)\phi(x_2)\phi(x_3)\phi(x_4)\rangle_{flat}=\frac{g(u,v)}{x_{12}^{2\Delta}{x_{34}^{2\Delta}}}
\end{equation},

g(u,v) can be computed using the procedure outlined in the appendix of [7]

Once the correlation functions are computed they can be integrated as followed to find the magnetization density. 

\begin{equation}
\langle\sigma^2\rangle= (\frac{1}{2 \pi^2 R^3})^2 \int dS_1 dS_2 \langle\phi(x_1)\phi(x_2)\rangle_{g_{uv}} 
\end{equation}
and

\begin{equation}
\langle\sigma^4\rangle= (\frac{1}{2 \pi^2 R^3})^4 \int dS_1 dS_2 dS_3 dS_4 \langle\phi(x_1)\phi(x_2)\phi(x_3)\phi(x_4)\rangle_{g_{uv}}
\end{equation}

The measure for these integrals are $dS_i=R^3\sin^2\psi_i \sin\theta_i d\psi_i d\theta_i d\phi_i $. For the rest of this paper because we are working with the unit 3 sphere we set R=1. The key for efficiently evaluating these integral is to use the SO(4) symmetry of $S^3$. For the two point function we evaluated over the following two arbitrary points on the surface of $S^3$ $ (\psi_1, \theta_1, \phi_1)$ and $(\psi_2, \theta_2, \phi_2)$. Using SO(4) rotations the reader can convince themselves that the first point can be set to $(0,0,0)$ while the second point is $(\psi_2,0,0)$. This results in the following integral for the 2nd order magnetization density. 

\begin{equation}
\int_0^{\pi } \frac{\left(\left(2 \pi ^2\right) (4 \pi ) \sin ^2\left(\psi
   _2\right)\right) \left(\frac{1}{2 \pi ^2}\right)^2}{\left(2 \cos
   ^2\left(\frac{0}{2}\right) 2 \cos ^2\left(\frac{\psi _2}{2}\right)\right){}^{0.518149}
   \left(\frac{\sin ^2\left(\psi _2\right)}{\left(1+\cos \left(\psi
   _2\right)\right){}^2}\right){}^{0.518149}} \, d\psi _2
\end{equation}

which yields from Mathematica's INTEGRATE function $\langle \sigma^2\rangle=0.84736$.

The four point function has the following coordinates $(\psi_1, \theta_1, \phi_1), (\psi_2, \theta_2, \phi_2), (\psi_3, \theta_3, \phi_3),  (\psi_4, \theta_4, \phi_4)$ . Using SO(4) we can simplfy the coordinates as follows $ (0,0,0),(\psi_2,0,0),(\psi_3, \theta_3 ,0), (\psi_4, \theta_4, \phi_4)$ The actual integral can be computed the same way $\langle\sigma^2\rangle$ was done, the only extra step is to evaluate the g(u,v) term   and write the conformal cross ratios, u and v in terms of the angles [3] which parametrize the 3 sphere. Using 15,000 Monte Carlo evaluations we calculate $\langle\sigma^4\rangle$ and quote the standard error as 1.59158(13) 

We now have all that we need to compute the Binder cumulant. 

\begin{equation}
U_4= \frac{3}{2} (1-\frac{1}{3} \frac{\langle\sigma^4\rangle}{\langle\sigma^2\rangle^2})
\end{equation}
which is $U_4$=0.3916843(910).

As a check that our procedure for evaluating the four-point function on the 3- sphere is correct we calculated the Binder cumulant for the free theory. The correlation functions for a free CFT are given in [14]. For a free CFT, the Binder cumulant should be zero. Using Monte Carlo integration over 1000 iterations, we calculated $U_4$=0.00002(48) . Our result is very comfortably within the range of the expected result of 0 for the Binder cumulant. We hope to check our results for the Binder cumulant of the critical 3D Ising model on $S^3$ using the QEM method in the near future.

\section{\label{sec:level1}Conclusion}

As shown with the simple forms of the correlation functions conformal invariance is a powerful tool which can be used to understand critical behavior. We have furthered the work of those who applied it to cases of curved two-dimensional geometries and flat 3 dimension space by calculating  the Binder cumulant of the critical 3D Ising model on the surface of a 3 sphere. We hope to match this result, which was done via tried and trued methods for calculating Binder cumulants to what the QEM predicts. If the QEM can successfully reproduce these results, then further work can be carried out to test more exotic geometries. Once the method is sufficiently demonstrated it can then be used for exciting applications such as particle creation during inflation in the strong field limit, non perturbative quantum effects at the event horizon of black holes and for analyzing a facet of fascinating systems

\section{ACKNOWLEDGMENTS}

I greatly thank George Fleming(Yale University) for advising me on this work and for guiding me in the right direction.This research was supported by Yale University 

\section{References}
[1] J. Penedones, E. Trevisani, and M. Yamazaki, Journal of High Energy Physics 2016, 70 (2016).

[2] M. Hogervorst, Journal of High Energy Physics 2016, 17 (2016).

[3] F. Kos, D. Poland, and D. Simmons-Duffin, Journal of High Energy Physics 2014, 91 (2014).

[4] F. Kos, D. Poland, and D. Simmons-Duffin, Journal of High Energy Physics 2014, 109 (2014).

[5] S. El-Showk, M. F. Paulos, D. Poland, S. Rychkov, D. Simmons-Duffin, and A. Vichi, Physical Review D 86, 025022 (2012).

[6] Z. Komargodski and D. Simmons-Duffin, Journal of Physics A: Mathematical and Theoretical 50, 154001 (2017).

[7] S. Rychkov, D. Simmons-Duffin, and B. Zan, SciPost Physics 2, 001 (2017).

[8] Y. Deng and H. W. Blo ̈te, Physical Review E 67, 036107 (2003).

[9] R. C. Brower, M. Cheng, E. S. Weinberg, G. T. Fleming, A. D. Gasbarro, T. G. Raben, and C.-I. Tan, Physical Review D 98, 014502 (2018).

[10] N. H. Kuiper, Annals of Mathematics , 916 (1949).

[11] A. A. Garc ́ıa, F. W. Hehl, C. Heinicke, and A. Macias, Classical and Quantum Gravity 21, 1099 (2004).

[12] H. Tom, “Greater2: A mathematica package for general relativity ”.

[13] F. Kos, D. Poland, D. Simmons-Duffin, and A. Vichi, Journal of High Energy Physics 2016, 36 (2016).

[14] A. L. Guerrieri, A. C. Petkou, and C. Wen, Journal of High Energy Physics 2016, 19 (2016).

\end{document}